\documentclass[referee,pdflatex,sn-mathphys]{sn-jnl}

\jyear{2021}%

\theoremstyle{thmstyleone}%
\theoremstyle{thmstyletwo}%

\theoremstyle{thmstylethree}%

\begin{document}

\title[ ]{Sampling-accelerated First-principles Prediction of Phonon Scattering Rates for Converged Thermal Conductivity and Radiative Properties}


\author[1,2]{\fnm{Ziqi} \sur{Guo}}\email{gziqi@purdue.edu}

\author[1,2]{\fnm{Zherui} \sur{Han}}\email{zrhan@purdue.edu}
\author[1,2]{\fnm{Dudong} \sur{Feng}}\email{fengdudong@purdue.edu}
\author*[1,3]{\fnm{Guang} \sur{Lin}}\email{guanglin@purdue.edu}
\author*[1,2]{\fnm{Xiulin} \sur{Ruan}}\email{ruan@purdue.edu}

\affil*[1]{\orgdiv{Department of Mechanical Engineering}, \orgname{Purdue University}, \orgaddress{\city{West Lafayette}, \postcode{47907-2088}, \state{IN}, \country{USA}}}

\affil[2]{\orgdiv{The Birck Nanotechnology Center}, \orgname{Purdue University}, \orgaddress{\city{West Lafayette}, \postcode{47907-2088}, \state{IN}, \country{USA}}}

\affil[3]{\orgdiv{Department of Mathematics}, \orgname{Purdue University}, \orgaddress{\city{West Lafayette}, \postcode{47907-2088}, \state{IN}, \country{USA}}}


\abstract{

First-principles prediction of thermal conductivity and radiative properties is crucial. However, computing phonon scattering, especially for four-phonon scattering, could be prohibitively expensive, and the thermal conductivity even for silicon was still under-predicted and not converged in the literature. Here we propose a method to estimate scattering rates from a small sample of scattering processes using maximum likelihood estimation. The computational cost of estimating scattering rates and associated thermal conductivity and radiative properties is dramatically reduced by over 99\%. \textcolor{black}{This allows us to use an unprecedented \textbf{q}-mesh of 32$\times$32$\times$32 for silicon and achieve a converged thermal conductivity value that agrees much better with experiments.} The accuracy and efficiency of our approach make it ideal for the high-throughput screening of materials for thermal and optical applications.

}




\maketitle
\pagebreak

\section{Introduction}\label{sec1}
Thermal conductivity and radiative properties are important material properties. Lattice thermal conductivity ($\kappa_{\rm l}$) is a key parameter for thermal management~\citep{moore2014emerging} and thermoelectrics~\citep{zebarjadi2012perspectives}. Thermal radiative properties, represented by the dielectric function ($\varepsilon$), are essential to applications including polaritonics~\citep{caldwell2015low}, thermal-photonic devices~\citep{feng2021near}, radiative energy converters~\citep{hu2018feature,feng2022thermoradiative} and radiative cooling~\citep{tong2022electronic}. Both of these properties are related to phonon-phonon scattering, the process by which atom vibrations in the material interact with each other. First principles perturbation theory can be used to predict the phonon scattering rates (${\tau_{\lambda}^{-1}}$) and subsequently these two properties~\citep{peierls1929kinetischen,ziman2001electrons,maradudin1962scattering, broido2007intrinsic,debernardi1995anharmonic}. While the phonon-phonon interaction up to the lowest order, i.e., three-phonon (3ph) scattering, had long been considered to be adequate for the prediction of ${\tau_{\lambda}^{-1}}$~\citep{broido2007intrinsic,esfarjani2011heat,lindsay2013first,tong2018temperature}, the importance of four-phonon (4ph) scattering, a higher order intrinsic scattering,  was recently revealed by Feng et al.~\citep{feng2016quantum, feng2017four}, which was confirmed by experiments for both lattice thermal conductivity~\citep{kang2018experimental,tian2018unusual,li2018high}, Raman linewidth~\citep{han2022raman} and infrared (IR) spectra~\citep{yang2020observation}. \textcolor{black}{Subsequent studies have shown that 4ph scattering is substantial in a wide range of materials~\citep{xia2018revisiting,ravichandran2018unified, Ashis2021Ultrahigh, jain2020multichannel, xia2020particlelike}. }\textcolor{black}{For accurate predictions of $\kappa_{\rm l}$ and $\varepsilon$, especially when dealing with materials exhibiting ultra-high or ultra-low thermal conductivity, or when operating at high temperatures, or investigating optical phonons, both three-phonon and four-phonon (3ph+4ph) scattering mechanisms should be assessed.}

However, the calculation of phonon scattering requires significant computational resources, especially for 4ph scattering. As a result, the first-principles predictions of $\kappa_{\rm l}$ and $\varepsilon$ have only been done for a limited number of materials. For various important materials used in solar cells, thermal barrier coatings, and thermoelectric devices, the complex crystal structures can lead to complex phonon dispersion and numerous phonon branches. This complexity can result in billions of phonon scattering processes, making 4ph or even 3ph scattering computationally infeasible.

In this paper, we propose an approach based on sampling and maximum likelihood estimation (MLE) to reduce the computational cost of phonon scattering calculations and accelerate the predictions of $\kappa_{\rm l}$ and $\varepsilon$. Under the relaxation time approximation (RTA), the sampling method aims to estimate the scattering rates of 3ph and 4ph scattering for each phonon mode $\lambda$ (denoted as ${\tau_{\lambda,\rm{3ph}}^{-1}}$ and ${\tau_{\lambda,\rm{4ph}}^{-1}}$, respectively) by sampling a subset from all phonon scattering processes. The concept is simple but works surprisingly well. After that, $\kappa_{\rm l}$ and $\varepsilon$ are determined by using the scattering rate of all phonon modes and the IR-active phonon mode only, respectively. We demonstrate that the sampling method can significantly reduce the computational cost of the first-principles predictions of thermal conductivity and radiative properties without sacrificing accuracy. \textcolor{black}{This allows us to revisit the thermal conductivity prediction of Si with an unprecedented \textbf{q}-mesh of 32$\times$32$\times$32, resulting in a converged thermal conductivity value that closely aligns with experimental data.} The accuracy and efficiency of our approach make it ideal for high-throughput screenings of materials for thermal and optical applications.

\section{Results}\label{sec2}
\subsection{Overview of the sampling method}

We start with an introduction to the rigorous calculation of ${\tau_{\lambda}^{-1}}$ and how we came up with the idea of using MLE to estimate it. The rigorous calculation of ${\tau_{\lambda,\rm{3ph}}^{-1}}$ and ${\tau_{\lambda,\rm{4ph}}^{-1}}$ requires an exhaustive calculation of the scattering rates of all possible 3ph and 4ph scattering processes for each phonon mode $\lambda$, denoted as $\Gamma_{\lambda \lambda^{\prime} \lambda^{\prime \prime}}^{\rm 3ph}$ and $\Gamma_{\lambda \lambda^{\prime} \lambda^{\prime \prime} \lambda^{\prime \prime \prime}}^{\rm 4ph}$, respectively. Then, under RTA, $\Gamma_{\lambda \lambda^{\prime} \lambda^{\prime \prime}}^{\rm 3ph}$ and $\Gamma_{\lambda \lambda^{\prime} \lambda^{\prime \prime} \lambda^{\prime \prime \prime}}^{\rm 4ph}$ are summed up separately to obtain ${\tau_{\lambda,\rm{3ph}}^{-1}}$ and ${\tau_{\lambda,\rm{4ph}}^{-1}}$ for phonon mode $\lambda$. To account for symmetry, multiple counting needs to be removed from the summation process (described in Method Section). At last, based on spectral Matthiessen's rule~\citep{ziman2001electrons}, we calculate the total phonon scattering rate: $\tau_{\lambda}^{-1}={\tau_{\lambda,\rm{3ph}}^{-1}}+{\tau_{\lambda,\rm{4ph}}^{-1}}$. Detailed formulations of 3ph and 4ph scattering rates can be found in~\cite{maradudin1962scattering,feng2016quantum}.

The huge computational cost of calculating $\tau_{\lambda}^{-1}$ originates from the substantial number of scattering processes involved. Calculating $\Gamma_{\lambda \lambda^{\prime} \lambda^{\prime \prime}}^{\rm 3ph}$ and $\Gamma_{\lambda \lambda^{\prime} \lambda^{\prime \prime} \lambda^{\prime \prime \prime}}^{\rm 4ph}$ of all of them can be expensive or even unaffordable, especially for 4ph scattering. To mitigate this computational cost, we start to consider the possibility of reducing the number of processes that we calculate. As ${\tau_{\lambda,\rm{3ph}}^{-1}}$ and ${\tau_{\lambda,\rm{4ph}}^{-1}}$ can be considered as the population sums of $\Gamma_{\lambda \lambda^{\prime} \lambda^{\prime \prime}}^{\rm 3ph}$ and $\Gamma_{\lambda \lambda^{\prime} \lambda^{\prime \prime} \lambda^{\prime \prime \prime}}^{\rm 4ph}$ under RTA, we can estimate the summations using MLE based on a randomly selected subset of scattering processes. ${\hat\tau_{\lambda,\rm{3ph}}^{-1}}$ and ${\hat\tau_{\lambda,\rm{4ph}}^{-1}}$ ( `$\hat{\ }$' is used to denote estimated values) are calculated by multiplying the average values of $\Gamma_{\lambda \lambda^{\prime} \lambda^{\prime \prime}}^{\rm 3ph}$ and $\Gamma_{\lambda \lambda^{\prime} \lambda^{\prime \prime} \lambda^{\prime \prime \prime}}^{\rm 4ph}$ for scattering processes in the subsets and the total number of scattering processes for each phonon mode $\lambda$. In this way, only the $\Gamma_{\lambda \lambda^{\prime} \lambda^{\prime \prime}}^{\rm 3ph}$ and $\Gamma_{\lambda \lambda^{\prime} \lambda^{\prime \prime} \lambda^{\prime \prime \prime}}^{\rm 4ph}$ of the small subset are calculated, resulting in significant savings of computational cost. The uncertainty of our estimation can be evaluated by calculating the confidence interval using statistical information of our sample. After calculating ${\hat\tau_{\lambda,\rm{3ph}}^{-1}}$ and ${\hat\tau_{\lambda,\rm{4ph}}^{-1}}$, we can use them to calculate $\hat\kappa_{\rm l}$ and $\hat\varepsilon$. The derivation of the maximum likelihood estimator of the population sum is shown in the Supplementary material. A detailed explanation of the workflow can be found in the Method Section.

\subsection{Maximum likelihood estimation of ${\tau_{\lambda}^{-1}}$ }

\begin{figure}[h]%
\centering
\includegraphics{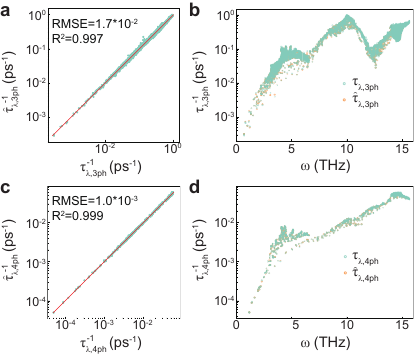}
\caption{\textbf{Maximum likelihood estimation of ${\tau_{\lambda,\rm{3ph}}^{-1}}$ and ${\tau_{\lambda,\rm{4ph}}^{-1}}$. } \textbf{a,} The scatter plot of ${\hat\tau_{\lambda,\rm{3ph}}^{-1}}$ with respect to ${\tau_{\lambda,\rm{3ph}}^{-1}}$. \textbf{b,} ${\tau}_{\lambda,{\rm 3ph}}^{-1}$ and ${\hat\tau}_{\lambda,\rm{3ph}}^{-1}$ as a function of phonon frequency. \textbf{c,} The scatter plot of ${\hat\tau_{\lambda,\rm{4ph}}^{-1}}$ with respect to ${\tau_{\lambda,\rm{4ph}}^{-1}}$. \textbf{d,} ${\tau}_{\lambda,{\rm 4ph}}^{-1}$ and ${\hat\tau}_{\lambda,\rm{4ph}}^{-1}$ as a function of phonon frequency. The estimation is for all phonon modes of Si at 300K, with sample size $5\times10^4$ and $5\times10^5$ for 3ph and 4ph, respectively. \textcolor{black}{The reported $R^2$ value is calculated from scattering rates in log scale.} }\label{Fig1}
\end{figure}

We take Si as an example to show the accuracy of our approach. Figure~\ref{Fig1} shows the model performance at sample size $n = 5\times10^4$ and $n = 5\times10^5$ for 3ph and 4ph, respectively. \textcolor{black}{Since the values of ${\tau_{\lambda,\rm{3ph}}^{-1}}$ and ${\tau_{\lambda,\rm{4ph}}^{-1}}$ vary orders of magnitude, the figures and the calculated $R^2$ values are based on scattering rates in logarithmic scale, while the result in normal scale is shown in Supplementary Fig.~1.} The high $R^2$ values indicate that our estimation of ${\tau_{\lambda,\rm{3ph}}^{-1}}$ and ${\tau_{\lambda,\rm{4ph}}^{-1}}$ are quite accurate. Additionally, Figure~\ref{Fig1}b and d show the consistency of estimation and rigorous results across the frequency range, as well as the satisfaction of the physical scaling law $\lim_{\omega \to 0} \tau_{\lambda}^{-1}=0$ for both 3ph and 4ph, which further supports the accuracy of our estimation. Note that the average sample sizes we use for 3ph and 4ph scattering represent only 6.25\% and 0.013\% of the total 3ph and 4ph scattering processes for each mode, respectively. This implies that our model is expected to yield significant time savings, as we will discuss in the subsequent section.

With the increase in sample size, the estimation accuracy is improved (Supplementary Fig.~2). This prompts two natural questions: What level of precision can we achieve with a given sample size? What is the appropriate sample size that yields an estimation with an acceptable level of uncertainty? While it's possible to determine the uncertainty through multiple random samplings and estimations, it conflicts with our goal of saving computational time. Consequently, it is preferable for us to obtain the uncertainty of our estimation from a single run.

\begin{figure}[h]%
\centering
\centerline{\includegraphics{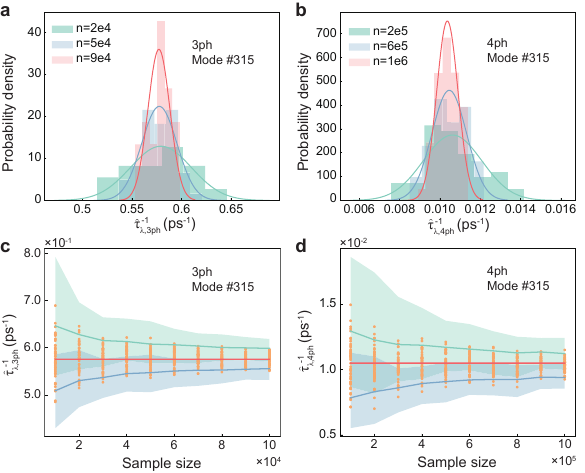}}
\caption{\textbf{Calculating confidence interval of ${\hat\tau_{\lambda,\rm{3ph}}^{-1}}$ and ${\hat\tau_{\lambda,\rm{4ph}}^{-1}}$. } All the results of ${\hat\tau_{\lambda,\rm{3ph}}^{-1}}$ and ${\hat\tau_{\lambda,\rm{4ph}}^{-1}}$ in this figure are for phonon mode \#315 at (0.5, 0, 0.5) of Brillouin zone and on the 3th phonon branch, with a phonon frequency of 9.62~THz. We perform random sampling and estimations 50 times for each sample size. \textbf{a, b,} The probability density of ${\hat\tau_{\lambda,\rm{3ph}}^{-1}}$ and ${\hat\tau_{\lambda,\rm{4ph}}^{-1}}$ with different sample sizes. \textbf{c, d,} The confidence interval for ${\hat\tau_{\lambda,\rm{3ph}}^{-1}}$ and ${\hat\tau_{\lambda,\rm{4ph}}^{-1}}$ with respect to sample sizes, respectively. The green and blue shaded areas are the upper and lower bound of confidence intervals and the green and blue lines are the means of the corresponding bounds, respectively. The orange dots are the estimation with different samples. The red lines are for the rigorously calculated scattering rate.}\label{Fig2}
\end{figure}

We take a phonon mode \#315 as an example to illustrate how we evaluate the uncertainty, where the index of phonon is defined within ShengBTE~\citep{li2014shengbte}. The position of this phonon mode on the phonon dispersion curve is shown in Supplementary Fig.~3. According to the Central Limit Theorem, when we sample a sufficiently large number of phonon scattering processes, ${\hat\tau_{\lambda,\rm{3ph}}^{-1}}$ and ${\hat\tau_{\lambda,\rm{4ph}}^{-1}}$ are approximately following normal distributions with means close to ${\tau}_{\lambda,{\rm 3ph}}^{-1}$ and ${\tau}_{\lambda,{\rm 4ph}}^{-1}$, and variances decreasing as sample sizes increase (Fig~\ref{Fig2}a and b). Based on this distribution, we can derive a confidence interval that specifies where the true values of ${\tau_{\lambda,\rm{3ph}}^{-1}}$ and ${\tau_{\lambda,\rm{4ph}}^{-1}}$ fall with a certain confidence level ($(1-\alpha)100\%$), serving as a measure of the estimation's uncertainty. \textcolor{black}{Since each scattering process is independently and randomly sampled from the underlying distribution of $\Gamma_{\lambda \lambda^{\prime} \lambda^{\prime \prime}}^{\rm 3ph}$ or $\Gamma_{\lambda \lambda^{\prime} \lambda^{\prime \prime} \lambda^{\prime \prime \prime}}^{\rm 4ph}$ of phonon mode $\lambda$, our method satisfies the assumption of the Central Limit Theorem that each random variable should be independent and identically distributed (i.i.d.).} We provide a rigorous derivation of the confidence interval for ${\tau_{\lambda,\rm{3ph}}^{-1}}$ and ${\tau_{\lambda,\rm{4ph}}^{-1}}$ in Method Section.

We select a confidence level of 90\%~($\alpha=0.1$) to calculate the confidence interval for each sample size. Then we perform 50 rounds of random samplings and estimations, and compare them with the confidence interval to verify the accuracy of our estimation (Fig.~\ref{Fig2}c and d). The results show that most of the estimated values fall within the confidence interval, indicating that our estimation is reliable. As the sample size increases, the upper and lower bounds of the confidence interval approach the true scattering rate, suggesting that the uncertainty decreases with more sampled phonon processes. With the confidence interval, we can determine whether the sample size is sufficiently large. The confidence interval goes below 10\% of the scattering rate at $n=20,000$ for ${\tau_{\lambda,\rm{3ph}}^{-1}}$ and $n=800,000$ for ${\tau_{\lambda,\rm{4ph}}^{-1}}$, which indicates that these sample sizes are sufficient for phonon mode \#315. We also verified our confidence interval on phonon mode \#6 and \#52, whose results are shown in Supplementary Fig.~4. Overall, the analytical derivation of the confidence interval serves to aid users in understanding the level of accuracy that can be achieved and in selecting an appropriate sample size.

\subsection{Time-saving in phonon scattering calculations}
\begin{figure}[h]%
\centering
\includegraphics{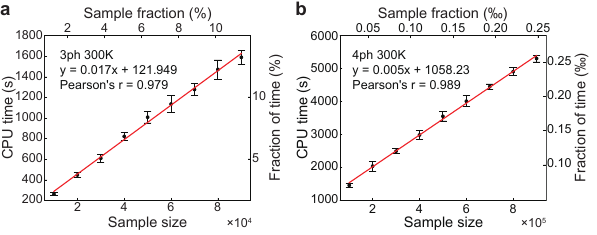}
\caption{\textbf{Average computational cost of estimating scattering rate with respect to sample sizes for Si.} \textbf{a,} ${\hat\tau_{\lambda,\rm{3ph}}^{-1}}$, \textbf{b,} ${\hat\tau_{\lambda,\rm{4ph}}^{-1}}$.  For each sample size, sampling and estimation are repeated 50 times, and the CPU time is represented with error bars based on one standard deviation. The upper x-axis displays the ratio of the sample size to the total number of phonon scattering processes, while the right y-axis displays the ratio of CPU time for the sampling method compared to the rigorous calculation.}\label{Fig3}
\end{figure}

Since we only rigorously compute a small subset of all the scattering processes, we anticipate significant computational cost savings. Figure~\ref{Fig3} and Supplementary~Fig.~5 illustrate the nearly linear relationship between the sample size and computational cost for Si and MgO, respectively. For phonon mode \#315 that we discussed in the previous section, we only need to calculate 2.5\% of all 3ph scattering processes to achieve an accurate ${\hat\tau_{\lambda,\rm{3ph}}^{-1}}$. This would lead to a 90.6\% reduction in computational cost compared to a full calculation. Comparing with ${\hat\tau_{\lambda,\rm{3ph}}^{-1}}$, ${\hat\tau_{\lambda,\rm{4ph}}^{-1}}$ requires an even smaller fraction of the sample (0.022\% of all 4ph scattering processes), leading to a significant reduction in computational cost (99.97\%). These substantial computational savings will significantly accelerate the calculation of $\hat\kappa_{\rm l}$ and $\hat\varepsilon$, which we shall now discuss in the following two sections.

\subsection{The estimation of lattice thermal conductivity }

To predict $\hat\kappa_{\rm{l,3ph+4ph}}$, we first obtain samples of 3ph and 4ph scattering processes for each phonon mode and calculate ${\hat\tau_{\lambda,\rm{3ph}}^{-1}}$ and ${\hat\tau_{\lambda,\rm{4ph}}^{-1}}$. ${\tau}_{\lambda}^{-1}$ is then determined using the spectral Matthiessen's rule~\citep{ziman2001electrons}: $\tau_{\lambda}^{-1}={\tau_{\lambda,\rm{3ph}}^{-1}}+{\tau_{\lambda,\rm{4ph}}^{-1}}$. Finally, we calculate $\hat\kappa_{\rm{l,3ph+4ph}}$ considering the spectral contribution of every phonon mode. When obtaining $\hat\kappa_{\rm{l,3ph}}$, only 3ph scattering is considered and ${\tau}_{\lambda}^{-1}$ contains only ${\tau}_{\lambda,\rm{3ph}}^{-1}$ term.

\begin{figure}[h]%
\centering
\centerline{\includegraphics[width=1.1\textwidth]{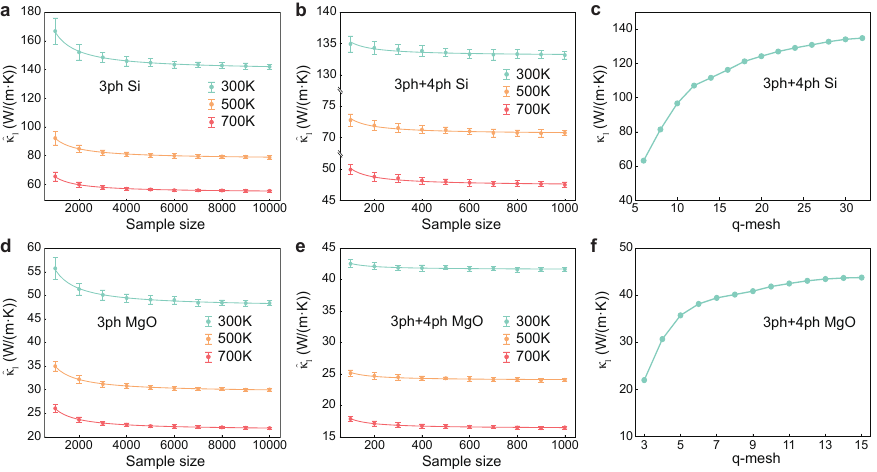}}
\caption{\textbf{Estimation of $\kappa_{\rm l}$ with the sampling method.}  Figures \textbf{a} and \textbf{d} show $\hat\kappa_l^{3ph}$ and figures \textbf{b} and \textbf{e} show $\hat\kappa_l^{3ph+4ph}$ as functions of sample size. Subplots \textbf{a} and \textbf{b} are for Si and subplots \textbf{d} and \textbf{e} are for MgO. For each sample size, sampling and estimation are repeated 50 times and the error bar is based on one standard deviation. Figures \textbf{c} and \textbf{f} show the convergence test of $\hat\kappa_l^{3ph+4ph}$ of Si and MgO at 300~K, respectively. 
}\label{Fig4}
\end{figure}

Figure~\ref{Fig4}a,b,d,e shows the relation between the sample size for each mode and $\hat\kappa_{\rm l}$ for Si and MgO. It is worth noting that for $\hat\kappa_{\rm{l,3ph+4ph}}$, the 3ph term ${\tau}_{\lambda,\rm{3ph}}^{-1}$ is calculated rigorously. Only the 4ph contribution ${\tau}_{\lambda,\rm{4ph}}^{-1}$ is estimated with our sampling method. This is to show the error and the saving of computational cost solely brought by the estimation of ${\tau}_{\lambda,\rm{4ph}}^{-1}$.
As the sample size increases, the uncertainty of estimation decreases, and the mean of $\hat\kappa_{\rm l}$ gradually approaches the true $\kappa_{\rm l}$ with a convergence rate proportional to $\frac{1}{n}$ for all temperatures and materials. The converged sample sizes are $9\times10^3$ for $\hat\kappa_{\rm{l,3ph}}$ and $8\times10^2$ for $\hat\kappa_{\rm{l,3ph+4ph}}$ for both materials, where the relative error for estimations with two consecutive sample sizes goes below 10\%. Notice that $\hat\kappa_{\rm{l,3ph}}$ and $\hat\kappa_{\rm{l,3ph+4ph}}$ tend to be higher than the corresponding true results, which is different from what we observe for ${\hat\tau_{\lambda,\rm{3ph}}^{-1}}$ and ${\hat\tau_{\lambda,\rm{4ph}}^{-1}}$ where the means of estimations remain close to the true result. This discrepancy arises because $\kappa_{\rm l}$ is inversely proportional to ${\tau}_{\lambda}^{-1}$, which amplifies the negative error and leads to an overestimation of $\hat\kappa_l$. Besides, we observe that $\hat\kappa_{\rm l}$ converges much faster than ${\hat\tau_{\lambda,\rm{3ph}}^{-1}}$ and ${\hat\tau_{\lambda,\rm{4ph}}^{-1}}$ as the sample size increases. The required sample sizes for reaching convergence are much smaller than those for estimating scattering rates. This is due to the existence of error-canceling effects when accumulating the spectral contribution of $\kappa_{\rm l}$ of all phonon modes. As a result, even if there are errors in ${\hat\tau_{\lambda,\rm{3ph}}^{-1}}$ and ${\hat\tau_{\lambda,\rm{4ph}}^{-1}}$, the final results of $\hat\kappa_{\rm l}$ would not be significantly impacted. Moreover, We find that for both Si and MgO, the converged sample size for 4ph is smaller than for 3ph. This is because both materials are primarily dominated by 3ph scattering, and the error of ${\hat\tau_{\lambda,\rm{4ph}}^{-1}}$ only contributes a small portion to the error of total thermal resistance, which suggests that a small sample size is enough to reach convergence for the estimation of $\kappa_{\rm{l,3ph+4ph}}$. \textcolor{black}{To further demonstrate the performance on materials with strong 4ph effects and different symmetry, we estimated $\kappa_{\rm{l,3ph}}$ and $\kappa_{\rm{l,3ph+4ph}}$ of LiCoO\textsubscript{2}, which is shown in Supplementary Fig.~6. }

\textcolor{black}{
When calculating the thermal conductivity, the first Brillouin zone should be discretized into \textbf{q}-mesh which should be determined based on a rigorous convergence test. However, due to the large computational cost, many studies use a small \textbf{q}-mesh for 4ph scattering calculation~\citep{xia2020high, zheng2022anharmonicity, chen2023strain}, which may not be converged. 
To reduce the computational cost, a smaller broadening factor (\textit{scaleboard} in ShengBTE~\cite{li2014shengbte}) was usually adopted in calculations. However, this method is known to underpredict the scattering rates. 
To get the result under high \textbf{q}-mesh, some meaningful attempts have been made by calculating thermal conductivity using several small \textbf{q}-meshes and extrapolating the result to the large \textbf{q}-mesh~\citep{li2014shengbte,gu2020thermal,gu2018thermal}. However, it lacks physical evidence to support a specific mathematical relationship between thermal conductivity and the \textbf{q}-mesh size. 
Since our method dramatically reduces the computational cost, it is now possible to study the 4ph scattering with a dense \textbf{q}-mesh. Take Si as an example. In previous studies, 
while the 3ph calculation is based on a dense \textbf{q}-mesh (around 28$\times$28$\times$28), the rigorous 3ph+4ph calculation can only be carried out on 16$\times$16$\times$16 \textbf{q}-mesh at most due to the limit of computational power~\citep{feng2016quantum, feng2017four, gu2020thermal, han2022fourphonon}. 
We do a convergence test on 3ph+4ph scattering at 300~K for Si using our sampling method, which is shown in Fig.~\ref{Fig4}c. The calculation of 3ph scattering rates is based on the iterative scheme and the calculation of 4ph scattering rates is based on RTA with a sample size $10^6$ to ensure the accuracy. The uncertainty of our method is only 0.2\% while the computational cost is dramatically reduced. We found that $\kappa_{\rm{l,3ph+4ph}}$ converged at 32$\times$32$\times$32 \textbf{q}-mesh (133 W/(m·K)), which is much higher than the result from 16$\times$16$\times$16 \textbf{q}-mesh (114 W/(m·K)) and is closer to the experimental results (130-150 W/(m·K)~\cite{morris1961thermal, shanks1963thermal,stuckes1960thermal, glassbrenner1964thermal}). For MgO, the 3ph+4ph thermal conductivity converges at 15$\times$15$\times$15 \textbf{q}-mesh (Fig.~\ref{Fig4}f), which is the same as the mesh used in previous study~\citep{han2023predictions}. 
In this paper, since we need to do a rigorous 3ph+4ph calculation to get a ground truth value to verify our estimations, the \textbf{q}-mesh is set to 16$\times$16$\times$16 for Si when illustrating the model performance of estimating ${\tau_{\lambda}^{-1}}$, the confidence interval and the time-saving. The estimation of thermal conductivity of Si that we have shown in Fig.~\ref{Fig4} is based on 32$\times$32$\times$32 \textbf{q}-mesh, while the result with 16$\times$16$\times$16 \textbf{q}-mesh is shown in Supplementary Fig.~7.}

\begin{figure}[h]%
\centering
\includegraphics{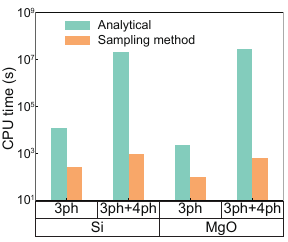}
\caption{\textbf{Time-saving of predicting lattice thermal conductivity.} The average CPU times for calculating $\hat\kappa_{\rm{l,3ph}}$ and $\hat\kappa_{\rm{l,3ph+4ph}}$ are based on 50 runs at their respective converged sample sizes of $9\times10^3$ and $8\times10^2$, respectively. The speedups are 47 times for Si with 3ph scattering (32$\times$32$\times$32 \textbf{q}-mesh), $2.3*10^4$ times for Si with 3ph+4ph scattering (16$\times$16$\times$16 \textbf{q}-mesh), 24 times for MgO with 3ph scattering and $4.4*10^4$ times for MgO with 3ph+4ph scattering (15$\times$15$\times$15 \textbf{q}-mesh).}\label{Fig5}
\end{figure}

To evaluate the time-saving effect, we test our method on Si and MgO at 300~K for both 3ph and 4ph calculations. For a sample size that reaches a relative uncertainty of less than 10\%, our method achieves four orders of magnitude acceleration for 4ph scattering calculations  (Fig.~\ref{Fig5}). The reduction in computational time is significant.

\subsection{The estimation of thermal radiative properties}
For the prediction of $\varepsilon$, ${\tau_{\lambda}^{-1}}$ of IR-active phonon modes are required, which are used as the damping factor ($\gamma$) in the Lorentz oscillator model to determine $\varepsilon$. Since Si has no IR-active phonon modes, we take MgO, a polar material, to demonstrate the performance of our method. 

\begin{figure}[h]%
\centering
\centerline{\includegraphics{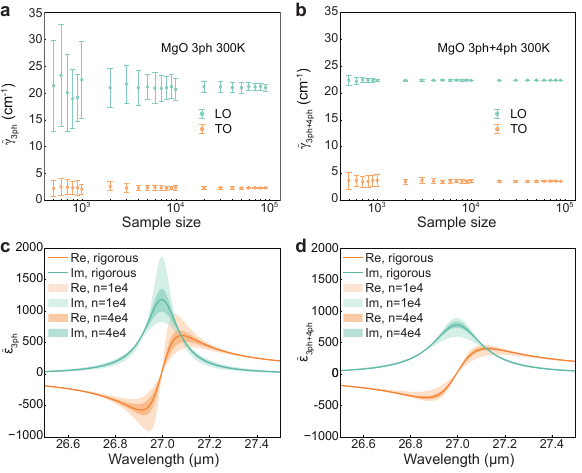}}
\caption{\textbf{Estimation of dielectric function for MgO at 300 K.} \textbf{a,} $\hat\gamma_{\mathrm{LO}, 3ph}$ and $\hat\gamma_{\mathrm{TO}, 3ph}$ with respect to different sample sizes. \textbf{b,} $\hat\gamma_{\mathrm{LO}, 3ph+4ph}$ and $\hat\gamma_{\mathrm{TO}, 3ph+4ph}$ with respect to different sample sizes. \textbf{c, d,} The dielectric function for 3ph ($\hat\varepsilon_{3ph}$) and 3ph+4ph ($\hat\varepsilon_{3ph+4ph}$), respectively. "Re" stands for the real part and "Im" stands for the imaginary part of the dielectric function, respectively. The rigorously calculated results are given by solid lines. The upper and lower boundaries of the shaded area are the maximum and minimum estimations of the 50 runs, so all estimations are within the shaded areas.
}\label{Fig6}
\end{figure}

Figure~\ref{Fig6}a and b show the relation between $\hat\gamma$ for both LO (longitudinal optical) and TO (transverse optical) modes and sample size. Again, for $\hat\gamma_{\mathrm{LO, 3ph+4ph}}$ and $\hat\gamma_{\mathrm{TO, 3ph+4ph}}$, the 3ph scattering part is calculated rigorously and 4ph scattering part is estimated with our sampling method. Similar to ${\hat\tau_{\lambda,\rm{3ph}}^{-1}}$ and ${\hat\tau_{\lambda,\rm{4ph}}^{-1}}$, increasing the sample size leads to a decrease in the uncertainty of damping factor. We then employ the Lorentz oscillator model with these damping factors to estimate $\hat\varepsilon_{\rm{3ph}}$ and $\hat\varepsilon_{\rm{3ph+4ph}}$, as shown in Figure~\ref{Fig6}c and d. To emphasize the uncertainty brought by the sampling method, we only show the wavelength range near the resonance frequency. Our sampling method gives more and more accurate estimations as the sample size increases. The predictions of a wider wavelength range and under different temperatures are shown in Supplementary Figs.~8-10, together with other thermal radiative properties including refractive index ($n$) and normal reflectance from a material-air interface ($R$). We take $6\times10^4$ for 3ph scattering and $2\times10^4$ for 4ph scattering as sufficiently large sample sizes where the relative error of the damping factors for both LO and TO modes go below 5\%. Regarding the time-saving effect, we can accelerate the calculation by four times and $3.4*10^4$ times for 3ph and 3ph+4ph scattering calculations, respectively.

\section{Discussion}

Our method is based on the RTA, which treats normal and umklapp scattering processes equally. This approximation is accurate for predicting thermal radiative properties because the optical responses only involve the excitation and decay of IR-active phonon modes, and Umklapp and normal processes are both effective decay channels for an excited (over-populated) zone-center optical phonon. Previous works have demonstrated the success of the RTA in predicting thermal radiative properties for such cases~\citep{yang2020observation,tong2022electronic}. For predicting lattice thermal conductivity, the RTA is suitable for materials in which the Umklapp processes are the dominant contributors to thermal resistance.

We should highlight that the sample is not from the phonon phase space but from all possible combinations of 3ph/4ph no matter whether they obey the energy or momentum conservation or not. Since the phonon phase space calculation accounts for nearly a quarter of the total computational time (Supplementary Fig.~11), if we first determine the phonon phase space and then perform sampling within it, the time-saving effect will be at most 78\%. Instead of taking this approach, we perform sampling based on all 3ph and 4ph  scattering processes before the evaluation of phonon phase space. Scattering rates of scattering processes that violate the energy or momentum selection rules are set to zero. By adopting this method, we can bypass the calculation of the phonon phase space on the entire phonon dispersion but effectively only do this for the chosen sample, hence achieving a time reduction of over 99\%. Furthermore, in a similar manner, we can also estimate the size of phonon phase space and the corresponding confidence interval using the sampling method, which is shown in the Supplementary material.

\textcolor{black}{
In our previous study~\cite{guo2023fast}, we introduced a machine learning-based approach that can reduce the computational cost of calculating phonon scattering by up to two orders of magnitudes. It's important to note that this machine learning method and the sampling method proposed in this paper are fundamentally different. The machine learning method in our previous work predicts the scattering rate of each individual scattering process ($\Gamma_{\lambda \lambda^{\prime} \lambda^{\prime \prime}}^{3ph}$ and $\Gamma_{\lambda \lambda^{\prime}  \lambda^{\prime \prime } \lambda^{\prime\prime\prime} }^{4ph}$) with a machine learning model trained on a subset of scattering processes. The scattering rate of one phonon mode (${\tau_{\lambda,\rm{3ph}}^{-1}}$ and ${\tau_{\lambda,\rm{4ph}}^{-1}}$) is calculated subsequently. Theoretically, this method can work both under RTA and with the iterative scheme, since it retains all the details of phonon scattering processes. On the other hand, the sampling method does not predict $\Gamma_{\lambda \lambda^{\prime} \lambda^{\prime \prime}}^{3ph}$ and $\Gamma_{\lambda \lambda^{\prime}  \lambda^{\prime \prime } \lambda^{\prime\prime\prime} }^{4ph}$, but directly estimate ${\tau_{\lambda,\rm{3ph}}^{-1}}$ and ${\tau_{\lambda,\rm{4ph}}^{-1}}$ based on the sample. This inherent simplicity enables the sampling method to achieve even greater computational efficiency, as it eliminates the need for a time-consuming training process. However, it does not retain detailed information about phonon scattering processes and is designed to operate exclusively under RTA. 
}

\textcolor{black}{
For the first-principles prediction of thermal conductivity, there are mainly two computationally expensive steps: calculating force constants and calculating phonon scattering. Our work aims to reduce the computational cost of the latter step, which by itself is a challenging task, while the former warrants further study in the future. There are some works that aim to accelerate the calculation of force constants such as using compressive sensing lattice dynamics~\citep{zhou2019compressive} and utilizing machine learning potentials to compute force constants~\citep{mortazavi2021accelerating, tang2023effect}. Future work could be done on combining our approach with these methods to reduce the total computational cost of predicting thermal conductivity.
}

There is some room to further improve our method. In our study, we employ sampling with replacement during the sampling process as it is computationally faster and more memory-efficient, given the large number of phonon scattering processes involved in our analysis. With sampling with replacement, the chosen scattering process is returned to the population after being selected, allowing for the possibility of selecting the same process multiple times during the sampling. However, theoretically, sampling without replacement can provide even greater accuracy \textcolor{black}{than sampling with replacement under the same sample size}. By removing each selected sample from the population after selection, sampling without replacement reduces the potential for redundancy in the sampled data, leading to a more diverse and representative sample. The confidence interval can be corrected by incorporating a finite population factor~\citep{thompson2012sampling} to account for the finite size of the population. \textcolor{black}{A detailed description is shown in the Supplementary Material.} Future work can focus on implementing this sampling scheme. Besides, when estimating $\kappa_{\rm l}$, we sample the same number of scattering processes from all phonon modes. However, since acoustic phonon modes have a larger contribution to $\kappa_{\rm l}$ compared to optical phonon modes, it would be more beneficial to focus on improving the accuracy of predictions for acoustic phonon modes. Thus, choosing different sample sizes for each mode based on their contributions to thermal conductivity can lead to a faster and more accurate estimation of $\kappa_{\rm l}$. The same thing happens to the estimation of thermal radiative properties, where the convergence rate of LO and TO modes can be quite different. The model efficiency can be further improved by choosing different sample sizes for these two modes.

To summarize, this study demonstrates a significant acceleration in the first-principles prediction of thermal conductivity and radiative properties, both of which are closely linked to phonon anharmonicity. By using a maximum likelihood estimation, we have successfully reduced the computational cost by over 99\%. This work removes the computational barrier associated with phonon scattering calculations, allowing for calculation with a converged \textbf{q}-mesh as well as high-throughput screening of materials' thermal and optical properties. 

\section{Method}

\subsection{Estimate ${\tau_{\lambda}^{-1}}$}

We first describe how we use MLE to estimate ${\tau_{\lambda,\rm{3ph}}^{-1}}$ and ${\tau_{\lambda,\rm{4ph}}^{-1}}$. The analytical equation for calculating ${\tau_{\lambda,\rm{3ph}}^{-1}}$ and ${\tau_{\lambda,\rm{4ph}}^{-1}}$ are:
\begin{align}
{\tau_{\lambda,\rm{3ph}}^{-1}}&={(\tau_{\lambda,\rm{3ph}}^{(+)})^{-1}} + {(\tau_{\lambda,\rm{3ph}}^{(-)})^{-1}} \text {, }\\
{\tau_{\lambda,\rm{4ph}}^{-1}}&={(\tau_{\lambda,\rm{4ph}}^{(++)})^{-1}}+{(\tau_{\lambda,\rm{4ph}}^{(+-)})^{-1}}+{(\tau_{\lambda,\rm{4ph}}^{(--)})^{-1}}\text {, }
\end{align}
where $(+)$ and $(-)$ terms stands for 3ph combination and splitting processes and the $(++)$, $(+-)$ and $(--)$ terms stand for 4ph combination, redistribution and splitting processes. Under RTA, the scattering rate of each type is given by~\citep{maradudin1962scattering,broido2007intrinsic,feng2016quantum,feng2017four}:

\begin{align}
{(\tau_{\lambda,\rm{3ph}}^{(+)})^{-1}} &=\frac{1}{N_{\rm{grid}}}\left(\sum_{\lambda^{\prime} \lambda^{\prime \prime}}^{(+)} \Gamma_{\lambda \lambda^{\prime} \lambda^{\prime \prime}}^{(+)}\right)\\
{(\tau_{\lambda,\rm{3ph}}^{(-)})^{-1}}&=\frac{1}{N_{\rm{grid}}}\left(\sum_{\lambda^{\prime} \lambda^{\prime \prime}}^{(-)} \frac{1}{2} \Gamma_{\lambda \lambda^{\prime} \lambda^{\prime \prime}}^{(-)}\right)\\
(\tau_{\lambda,\rm{4ph}}^{(++)})^{-1}&=\frac{1}{N_{\rm{grid}}}\left(\sum_{\lambda^{\prime} \lambda^{\prime \prime} \lambda^{\prime \prime \prime}}^{(++)} \frac{1}{2} \Gamma_{\lambda \lambda^{\prime} \lambda^{\prime \prime} \lambda^{\prime \prime \prime}}^{(++)}\right) \text {, }\\
(\tau_{\lambda,\rm{4ph}}^{(+-)})^{-1}&=\frac{1}{N_{\rm{grid}}}\left(\sum_{\lambda^{\prime} \lambda^{\prime \prime}\lambda^{\prime \prime \prime}}^{(+-)} \frac{1}{2} \Gamma_{\lambda \lambda^{\prime} \lambda^{\prime \prime}\lambda^{\prime \prime \prime}}^{(+-)}\right) \text {, }\\
(\tau_{\lambda,\rm{4ph}}^{(--)})^{-1}&=\frac{1}{N_{\rm{grid}}}\left(\sum_{\lambda^{\prime} \lambda^{\prime \prime} \lambda^{\prime \prime \prime}}^{(--)} \frac{1}{6} \Gamma_{\lambda \lambda^{\prime} \lambda^{\prime \prime} \lambda^{\prime \prime \prime}}^{(--)}\right) \text {, }
\end{align}
where $N_{\rm{grid}}$ stands for the total grid of \textbf{q}-mesh and $\Gamma$ terms stand for the scattering rates of phonon scattering processes, which can be computed using Fermi's golden rule~\citep{dirac1927quantum} and has been described in previous literature. The fractions in the summation account for the multiple counting of phonon scattering processes originated from symmetry. When illustrating the model performance of estimating ${\tau_{\lambda}^{-1}}$, the confidence interval and the time saving, the \textbf{q}-mesh for Si is set to 32$\times$32$\times$32 for 3ph and 16$\times$16$\times$16 for 4ph scattering. For MgO, both 3ph and 4ph scattering are using a meth of 15$\times$15$\times$15. {When estimating $\kappa_{\rm l}$ and $\varepsilon$, the \textbf{q}-mesh is set to 32$\times$32$\times$32 for Si and 15$\times$15$\times$15 for MgO, respectively. } The broadening factor is kept at unity for all cases.

We sample $n_{\lambda}$ scattering processes from all phonon scattering processes of a phonon mode $\lambda$. The total number of phonon scattering processes $N_{\lambda}$ is given by: 
\begin{equation}
N_{\lambda}=\left\{
\begin{aligned}
N_{\rm{bands}}^2N_{\rm{grid}}  &  & \text{3ph scattering}\\
N_{\rm{bands}}^3N_{\rm{grid}}^2 &  & \text{4ph scattering}
\end{aligned}
\right.
\end{equation}
where $N_{\rm{bands}}$ is the number of phonon branches.

After sampling, we use the MLE to estimate ${\tau_{\lambda}^{-1}}$, which is given by: 

\begin{align}
{(\hat\tau_{\lambda,\rm{3ph}}^{(+)})^{-1}} &=\frac{1}{N_{\rm{grid}}}\left(\frac{N_{\lambda}^{(+)}}{n_{\lambda}^{(+)}} \sum_{\lambda_n^{\prime} \lambda_n^{\prime \prime}}^{(+)} \Gamma_{\lambda \lambda_n^{\prime} \lambda_n^{\prime \prime}}^{(+)}\right)\\
{(\hat\tau_{\lambda,\rm{3ph}}^{(-)})^{-1}}&=\frac{1}{N_{\rm{grid}}}\left(\frac{N_{\lambda}^{(-)}}{n_{\lambda}^{(-)}} \sum_{\lambda_n^{\prime} \lambda_n^{\prime \prime}}^{(-)} \frac{1}{2} \Gamma_{\lambda \lambda_n^{\prime} \lambda_n^{\prime \prime}}^{(-)}\right)\\
(\hat\tau_{\lambda,\rm{4ph}}^{(++)})^{-1}&=\frac{1}{N_{\rm{grid}}}\left(\frac{N_{\lambda}^{(++)}}{n_{\lambda}^{(++)}} \sum_{\lambda_n^{\prime} \lambda_n^{\prime \prime} \lambda_n^{\prime \prime \prime}}^{(++)} \frac{1}{2} \Gamma_{\lambda \lambda_n^{\prime} \lambda_n^{\prime \prime} \lambda_n^{\prime \prime \prime}}^{(++)}\right) \text {, }\\
(\hat\tau_{\lambda,\rm{4ph}}^{(+-)})^{-1}&=\frac{1}{N_{\rm{grid}}}\left(\frac{N_{\lambda}^{(+-)}}{n_{\lambda}^{(+-)}} \sum_{\lambda_n^{\prime} \lambda_n^{\prime \prime}\lambda_n^{\prime \prime \prime}}^{(+-)} \frac{1}{2} \Gamma_{\lambda \lambda_n^{\prime} \lambda_n^{\prime \prime}\lambda_n^{\prime \prime \prime}}^{(+-)}\right) \text {, }\\
(\hat\tau_{\lambda,\rm{4ph}}^{(--)})^{-1}&=\frac{1}{N_{\rm{grid}}}\left(\frac{N_{\lambda}^{(--)}}{n_{\lambda}^{(--)}} \sum_{\lambda_n^{\prime} \lambda_n^{\prime \prime} \lambda_n^{\prime \prime \prime}}^{(--)} \frac{1}{6} \Gamma_{\lambda \lambda_n^{\prime} \lambda_n^{\prime \prime} \lambda_n^{\prime \prime \prime}}^{(--)}\right) \text {, }
\end{align}
where $\lambda_n^{\prime} \lambda_n^{\prime \prime}\lambda_n^{\prime \prime \prime}$ and $\lambda_n^{\prime} \lambda_n^{\prime \prime}\lambda_n^{\prime \prime \prime}$ stand for the sampled 3ph and 4ph scattering process for phonon mode $\lambda$, $n_{\lambda}$ stands for sample size.

\subsection{Confidence interval of ${\tau_{\lambda}^{-1}}$}
We take $(\tau_{\lambda,\rm{3ph}}^{(+)})^{-1}$ as an example to show the derivation of the confidence interval. The mean $\bar\Gamma_{\lambda}^{(+)}$ and variance $(s_{\lambda}^{(+)})^2$ of the sampled $\Gamma_{\lambda \lambda_n^{\prime} \lambda_n^{\prime \prime}}^{(+)}$ is given by:
\begin{align}
\bar\Gamma_{\lambda}^{(+)} &= \frac{1}{n_{\lambda}^{(+)}}\sum_{\lambda_n^{\prime} \lambda_n^{\prime \prime}}^{(+)} \Gamma_{\lambda \lambda_n^{\prime} \lambda_n^{\prime \prime}}^{(+)} \\
(s_{\lambda}^{(+)})^2&=\sum_{\lambda_n^{\prime} \lambda_n^{\prime \prime}}^{(+)} \frac{\left(\Gamma_{\lambda \lambda_n^{\prime} \lambda_n^{\prime \prime}}^{(+)}-\bar\Gamma_{\lambda}^{(+)}\right)^2}{{n_{\lambda}^{(+)}}-1}
\end{align}

According to Central Limit Theorem, the distribution of the sample mean $\bar\Gamma_{\lambda}^{(+)}$ is approximately a normal distribution with a variance given by:
\begin{equation}
{Var}(\bar\Gamma_{\lambda}^{(+)}) =\frac{(s_{\lambda}^{(+)})^2}{{n_{\lambda}^{(+)}}}
\end{equation}

The variance of ${(\hat\tau_{\lambda,\rm{3ph}}^{(+)})^{-1}}$ is then given by:
\begin{equation}
{Var}({(\hat\tau_{\lambda,\rm{3ph}}^{(+)})^{-1}}) = {Var}(N_{\lambda}^{(+)}\bar\Gamma_{\lambda}^{(+)}) = N_{\lambda}^{(+)}\frac{s^2}{{n_{\lambda}^{(+)}}}
\end{equation}

As the sample variance is approximated, we use t-distribution to calculate the confidence interval of ${(\hat\tau_{\lambda,\rm{3ph}}^{(+)})^{-1}}$. The $(1-\alpha)100\%$ confidence interval for ${(\hat\tau_{\lambda,\rm{3ph}}^{(+)})^{-1}}$ is given by:
\begin{equation}
(\hat\tau_{\lambda,\rm{3ph}}^{(+)})^{-1} \pm {t_{\alpha/2,n_{\lambda}^{(+)}}} N_{\lambda}^{(+)} \sqrt{{{s_{\lambda}^{(+)}}^2}/{n_{\lambda}^{(+)}}} 
\end{equation}
where $t_{\alpha/2,n_{\lambda}^{(+)}}$ is the t-value of a t-distribution with $(n_{\lambda}^{(+)}-1)$ degrees of freedom and significant level $\alpha/2$. 

Similarly, we can derive the confidence interval of ${(\hat\tau_{\lambda,\rm{3ph}}^{(-)})^{-1}}$, which is given by:
\begin{equation}
(\hat\tau_{\lambda,\rm{3ph}}^{(-)})^{-1} \pm {t_{\alpha/2,n_{\lambda}^{(-)}}} N_{\lambda}^{(-)} \frac{1}{2} \sqrt{{{s_{\lambda}^{(-)}}^2}/{n_{\lambda}^{(-)}}} 
\end{equation}

We denote the half length of the confidence interval with the symbol $CI$ and we have:
\begin{align}
CI^{(+)}_{\lambda,\rm{3ph}} =  {t_{\alpha/2,n_{\lambda}^{(+)}}} N_{\lambda}^{(+)} \sqrt{{{s_{\lambda}^{(+)}}^2}/{n_{\lambda}^{(+)}}} \\
CI^{(-)}_{\lambda,\rm{3ph}} =  {t_{\alpha/2,n_{\lambda}^{(-)}}} N_{\lambda}^{(-)} \frac{1}{2}\sqrt{{{s_{\lambda}^{(-)}}^2}/{n_{\lambda}^{(-)}}} 
\end{align}

When summing up scattering rates in different categories, based on the rule of error propagation, the confidence interval of ${\hat\tau}_{\lambda,\rm{3ph}}^{-1}$ is given by:
\begin{align}
{\hat\tau}_{\lambda,\rm{3ph}}^{-1} \pm  \left(   \left( CI^{(+)}_{\lambda,\rm{3ph}}\right)^2 
  +   \left(CI^{(-)}_{\lambda,\rm{3ph}}\right)^2\right)^{1/2}
\end{align}

The confidence interval of ${\hat\tau}_{\lambda,\rm{4ph}}^{-1}$ can be derived in a similar manner: 
\begin{align}
{\hat\tau}_{\lambda,\rm{4ph}}^{-1} \pm  \left(   \left( CI^{(++)}_{\lambda,\rm{4ph}}\right)^2 
  + \left(CI^{(+-)}_{\lambda,\rm{4ph}}\right)^2 + \left(CI^{(--)}_{\lambda,\rm{4ph}}\right)^2 \right)^{1/2}
\end{align}
where $CI^{(++)}_{\lambda,\rm{4ph}}$, $CI^{(+-)}_{\lambda,\rm{4ph}}$ and $CI^{(--)}_{\lambda,\rm{4ph}}$ are given by: 
\begin{align}
CI^{(++)}_{\lambda,\rm{4ph}} &= {t_{\alpha/2,n_{\lambda}^{(++)}}} N_{\lambda}^{(++)} \frac{1}{2}\sqrt{{{s_{\lambda}^{(++)}}^2}/{n_{\lambda}^{(++)}}} \\
CI^{(+-)}_{\lambda,\rm{4ph}} &= {t_{\alpha/2,n_{\lambda}^{(+-)}}} N_{\lambda}^{(+-)} \frac{1}{2}\sqrt{{{s_{\lambda}^{(+-)}}^2}/{n_{\lambda}^{(+-)}}} \\
CI^{(--)}_{\lambda,\rm{4ph}} &= {t_{\alpha/2,n_{\lambda}^{(--)}}} N_{\lambda}^{(--)} \frac{1}{6}\sqrt{{{s_{\lambda}^{(--)}}^2}/{n_{\lambda}^{(--)}}}
\end{align}

\subsection{Predicting lattice thermal conductivity}

The total scattering rate of a phonon mode $\tau_{\lambda}$ is calculated based on spectral Matthiessen's rule~\citep{ziman2001electrons}: 

\begin{equation}
\tau_{\lambda}^{-1}=\left\{
\begin{aligned}
{\tau}_{\lambda,\rm{3ph}}^{-1} &  & \text{3ph scattering}\\
{\tau}_{\lambda,\rm{3ph}}^{-1}+{\tau}_{\lambda,\rm{4ph}}^{-1} &  & \text{3ph+4ph scattering}
\end{aligned}
\right.
\end{equation}

 The lattice thermal conductivity is calculated by:
\begin{equation}
\kappa_l=\frac{1}{V}\sum_{\lambda}{c_{\lambda}}v_{\lambda}^2\tau_{\lambda}\text {, }
\end{equation}
where $V$ is the unit cell volume, $v_{\lambda}$ and $c_{\lambda}$ are the group velocity and the specific heat of phonon mode $\lambda$, respectively.

\subsection{Predicting thermal radiative properties}

The complex dielectric function of polar dielectrics in the mid-IR range can be described by the Lorentz oscillator model~\citep{born1955dynamical, barker1964transverse,gervais1974anharmonicity}:

\begin{equation}
\varepsilon(\omega)=\varepsilon_{\infty}    \prod_m  \left(\frac{\omega_{\mathrm{LO}, m}^2-\omega^2-i \gamma_{\mathrm{LO},m} \omega }{\omega_{\mathrm{TO}, m}^2-\omega^2- i \gamma_{\mathrm{TO},m} \omega}\right)
\end{equation}
where $\varepsilon_{\infty}$ is the dielectric constant at the high-frequency limit, $\omega$ is the photon frequency, $\omega_{\mathrm{LO},m}$ and $\omega_{\mathrm{TO},m}$ are frequencies of the zone-center IR active LO an TO phonon modes, respectively. $\gamma_{\mathrm{LO},m}$ and $\gamma_{\mathrm{TO},m}$ are the damping factors, which can be derived from the phonon scattering rate of zone-center IR active LO and TO phonon modes, respectively.

From the dielectric function, we can further derive many useful thermal radiative properties including the complex refractive index ($m$) and the normal reflectance of the air-materials interface ($R$).
\begin{align}
m&=\sqrt{\varepsilon} \\
R&= \left\lvert \frac{\sqrt{\varepsilon}-1}{\sqrt{\varepsilon}+1} \right\rvert^2
\end{align}

\backmatter





\section*{Data availability}

The original results of the study are available from the corresponding authors upon reasonable request.

\section*{Acknowledgments}
The thermal radiative properties portion of the work was supported by National Science Foundation (Award No. 2102645). The thermal conductivity portion was supported by National Science Foundation (Award No. 2321301), and the methodology will be implemented into an open-source code FourPhonon and that effort was supported by National Science Foundation (Award No. 2311848). 


\section*{Code availability}
The work is incorporated as a new feature within the FourPhonon package and is available at~\href{https://github.com/FourPhonon/FourPhonon}{https://github.com/FourPhonon/FourPhonon}.


\section*{Competing interests}
The authors declare no competing interests.

\section*{Author contributions}
Z.G., X.R. and G.L. conceived the study. Z.G. designed and implemented the models, analyzed the results and wrote the manuscript. Z.H. and D.F. helped with the data analysis. X.R. and G.L. supervised the project. All authors contributed to discussions and revisions of the manuscript.

\bibliography{sn-bibliography}


\end{document}